\documentclass{elsart1p} 
 
 \usepackage{graphicx} 
 \usepackage{epsfig} 
 
\usepackage{amssymb} 
 
\begin{document} 
 
\begin{frontmatter} 
 
 
 
\title{Chiral dynamics of few-nucleon systems} 
 
 
\author{E.~Epelbaum} 
\ead{e.epelbaum@fz-juelich.de} 
\address{Forschungszentrum J\"ulich, Institut f\"ur Kernphysik, 
D-52425 J\"ulich, Germany} 

\address{Universit\"at Bonn, Helmholtz-Institut f{\"u}r
  Strahlen- und Kernphysik, D-53115 Bonn, Germany}
 
\begin{abstract} 
I discuss some recent developments in chiral effective field theory for
few-nucleon systems.  
\end{abstract} 
 
\end{frontmatter} 
 
\section{Introduction} 
\label{sec:intro} 
 
 

Almost two decades ago Weinberg formulated his ideas to extend chiral
perturbation theory to systems with several nucleons \cite{Weinberg:1990rz}. This seminal
work initiated extensive research towards better understanding of the structure
of nuclear forces and the dynamics of few- and many-nucleon systems based on the
powerful framework of effective field theory (EFT). In this talk I outline the
current status of chiral EFT for nuclear forces and applications in the
few-nucleon sector. I discuss the structure of the subleading contributions to the
three-nucleon force (3NF) and the leading four-nucleon force (4NF) which are
currently being investigated. The role of $\Delta$-excitation will also be
addressed. The last part of this talk concerns lattice simulations of light
nuclei.

\section{Chiral EFT for few nucleons: present status} 
\label{sec:status} 
  
Chiral EFT is an appropriate framework to analyze 
the properties of few-nucleon systems at low energy.\footnote{At very low
  energies, it is advantageous to use pionless EFT, see \cite{Braaten:2004rn} for a recent
  review article.} It is based on the
most general effective Lagrangian for Goldstone bosons (pions) and matter
fields (nucleons and perhaps deltas) consistent with the chiral symmetry of
QCD. For energies below the pion-production threshold, a nonrelativistic
treatment of the nucleons is justified. Further, it is advantageous to
eliminate the pionic degrees of freedom which gives
rise to the (nonlocal) nuclear Hamiltonian 
\begin{equation}
H_{\rm nucl} = H_0 + V_{2N} + V_{3N} + V_{4N} + \ldots\,.
\end{equation}
The resulting quantum-mechanical $A$-body problem can be dealt with using
rigorous few-body techniques or many-body methods. 

The derivation of the nuclear potentials from the effective
Lagrangian can be carried out straightforwardly utilizing the so-called chiral
expansion. The importance of a particular contribution to the nuclear
Hamiltonian is determined by the corresponding power of the expansion parameter
$Q/\Lambda$ where $Q$ and $\Lambda$ refer to the generic low-momentum scale associated with 
external nucleon momenta or $M_\pi$ and the pertinent hard scale, 
respectively \cite{Weinberg:1990rz}. For the two-nucleon force (2NF), this
expansion has the form
\begin{equation}
V_{2N} = V_{2N}^{(0)} + V_{2N}^{(2)} + V_{2N}^{(3)} + V_{2N}^{(4)} + \ldots \,,
\end{equation}
where the subscripts refer to the chiral order. 
The leading contribution $V_{2N}^{(0)}$ is due to $1\pi$-exchange (OPE)
and two NN contact interactions without derivatives.
The corrections result from contact interactions with increasing number of
derivatives and/or insertions of $M_\pi^2$ and exchange of pions. In
particular, $2\pi$-exchange starts to contribute at next-to-leading order
(NLO) and provides an important ingredient of the nuclear force
\cite{Rentmeester:1999vw}. At present, 
the 2NF has been worked out and applied in the NN system at
next-to-next-to-next-to-leading 
order (N$^3$LO) \cite{Entem:2003ft,Epelbaum:2004fk}, see \cite{Ordonez:1995rz} for the
pioneering work along this line. Its long-range part
involves contributions from $1\pi$- $2\pi$- and $3\pi$-exchange and is
parameter-free since the corresponding low-energy constants (LECs) are known
from $\pi N$ scattering. Numerically, the $3\pi$-exchange potential 
turns out to be negligibly small \cite{Kaiser:1999ff}.
The short-range part consists 
of 24 independent contact interactions\footnote{This number does not include
  isospin-breaking terms and refers to Weinberg's counting rules 
  based on the naive dimensional analysis. Alternative counting
  schemes are currently being explored.} whose strengths were determined
from fits to NN low-energy data, see Ref.~\cite{Epelbaum:2001fm} for the
discussion on resonance saturation of the corresponding LECs. Both 
N$^3$LO calculations of Ref.~\cite{Entem:2003ft} and \cite{Epelbaum:2004fk}
yield compatible results for the NN system (within the theoretical uncertainty)
and demonstrate an accurate description of the low-energy scattering data and
the deuteron properties. 

\begin{figure}[tb]
\vskip 1 true cm
  \begin{center} 
\includegraphics[width=\textwidth,keepaspectratio,angle=0,clip]{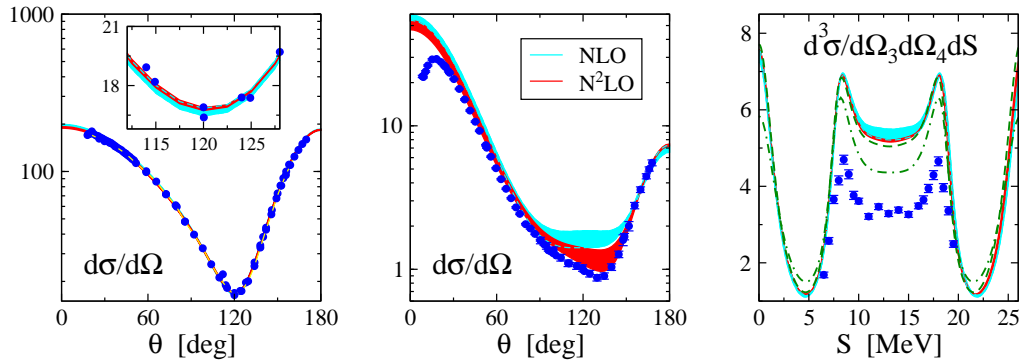}
    \caption{
         Differential cross section (in mb/sr) for
         elastic $Nd$ scattering at 10 MeV (left panel) and 65 MeV (middle
         panel) at NLO (light-shaded bands) and N$^2$LO (dark-shaded
         bands). References to data can be found in
         \cite{Epelbaum:2002vt}. Right panel shows $pd$ breakup cross section (in mb
         MeV$^{-1}$ sr$^{-2}$) along the kinematical locus S at 19 MeV nucleon
         energy in the SCRE configuration with $\alpha = 56^\circ$ \cite{Ley:2006hu}. Dashed and
         dashed-dotted lines are results based on the CD
         Bonn 2000 2NF \cite{Machleidt:2000ge} combined with the TM99 3NF
         \cite{Coon:2001pv} and the coupled  channel
         calculation including the explicit $\Delta$ and the Coulomb
         interaction \cite{Deltuva:2005cc}, respectively.
\label{fig1} 
 }
  \end{center}
\end{figure}

Systems with three- and more nucleons provide an excellent testing ground for
chiral nuclear forces and were extensively studied in this framework 
over the past few years. Chiral power counting explains naturally the
observed hierarchy of nuclear forces $V_{2N} \gg
V_{3N} \gg V_{4N} \ldots $ with  
\begin{equation}
V_{3N} =  V_{3N}^{(3)} + V_{3N}^{(4)} + \ldots \,, \quad  
V_{4N} =  V_{3N}^{(4)} + V_{3N}^{(5)} + \ldots \,,\quad  \ldots \,.
\end{equation}
The first nonvanishing contribution to the 3NF appears at N$^2$LO
relative to the leading 2NF and results from $2\pi$-exchange, $1\pi$-exchange
with 2N contact interaction and the purely short-range 3N contact term. While
the $2\pi$-exchange contribution is parameter-free, the two other topologies
depend on the unknown LECs $D$ and $E$ which were fitted to the $^3$H
binding energy and $nd$ doublet
scattering length \cite{Epelbaum:2002vt} and to the combinations of the $^3$H/$^4$He 
\cite{Nogga:2004aa} and $^3$H/$^{10}$B \cite{Navratil:2007we} binding energies.  
The resulting nuclear Hamiltonian was extensively tested in the 3N continuum
\cite{Ley:2006hu,Witala:2006nn,Kistryn:2005fi,Duweke:2004xv} and applied to study the
properties of light nuclei using  the no-core shell model approach 
\cite{Nogga:2004aa,Navratil:2007we,Nogga:2005hp}. The calculated 3N scattering
observables are generally in a reasonable agreement with the data which
improves when going from NLO to N$^2$LO, see e.g.~the left and middle panels 
in Fig.~\ref{fig1}. 
Notice however that the theoretical uncertainty at N$^2$LO appears to be
rather large already at moderate energies, see the middle panel in Fig.~\ref{fig1}. 
We further emphasize that there are certain observables for which  
remarkably large discrepancies with the data are found even at very low energy. A well-known
example is the so-called symmetric space-star (SST) $Nd$ breakup configuration 
at $E_{N} = 13$ MeV \cite{Epelbaum:2000mx,Epelbaum:2002vt}, in which the
plane in the c.m.~system spanned by the 
outgoing nucleons is perpendicular to the beam axis. Recently, $pd$ data for a
similar symmetric constant relative-energy configuration have been measured  
at $E_d = 19$ MeV \cite{Ley:2006hu}. This geometry is characterized by the angle $\alpha$
between the beam axis and the plane in the c.m.~system spanned by the 
outgoing nucleons. Similar to the SST
geometry, one observes large deviations between the theory and the data,
see right panel in Fig.~\ref{fig1},\footnote{We emphasize that chiral EFT results are obtained
  without taking into account Coulomb
  interaction.} which also hold for calculations based on
modern phenomenological potentials. The included 3NFs have
little effect on the cross section while the effect of the Coulomb interaction 
is significant but removes only a part of the discrepancy. For more details
on these and related topics the reader is referred to Ref.~\cite{Epelbaum:2005pn}, 
see also Ref.~\cite{Bedaque:2002mn}.

\section{N$^3$LO contributions to three- and four-nucleon forces} 
\label{sec:n3lo} 

\begin{figure}[tb]
\vskip 1 true cm
  \begin{center} 
\includegraphics[width=12.5cm,keepaspectratio,angle=0,clip]{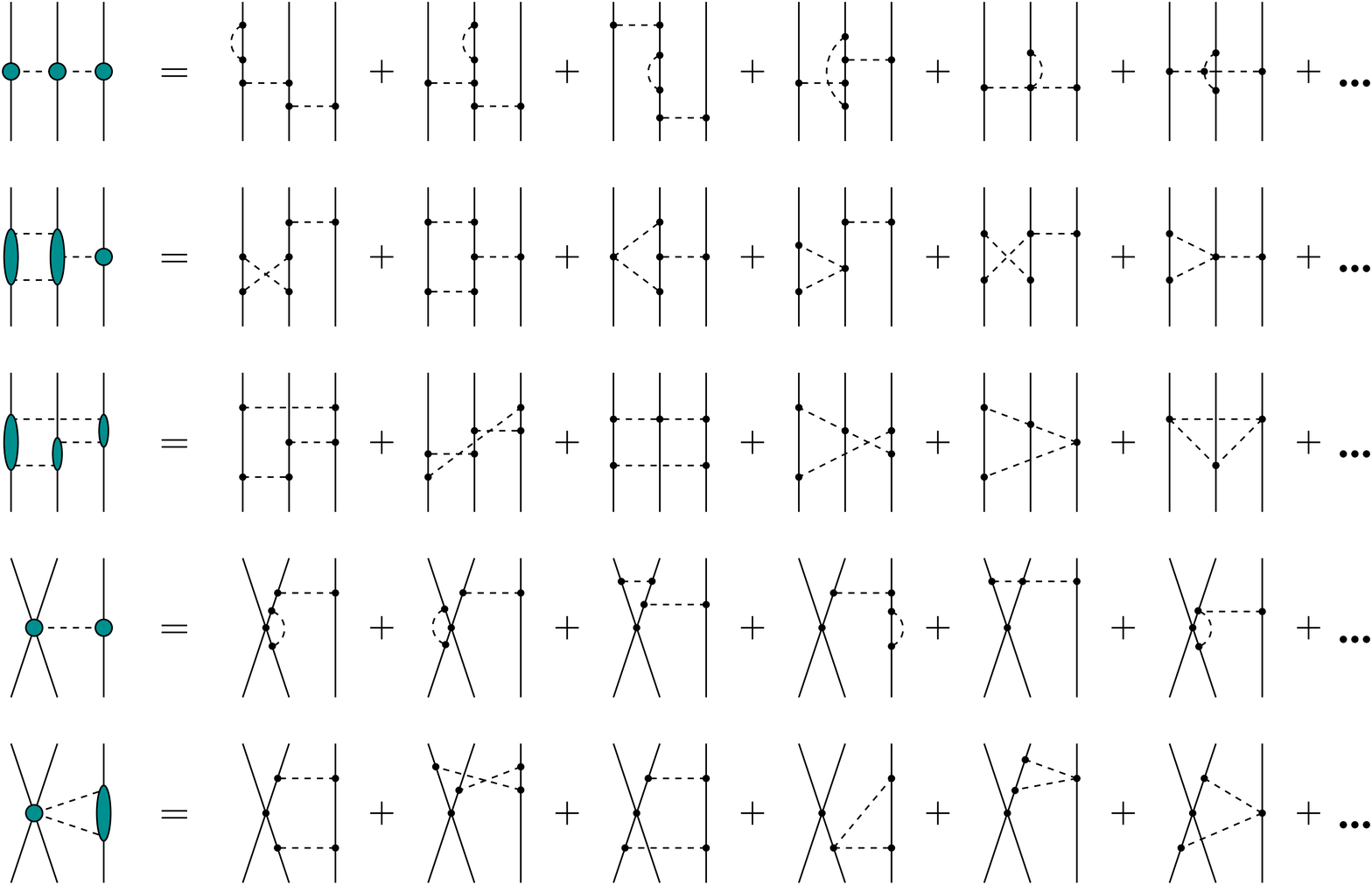}
    \caption{
         Different 3NF contributions at N$^3$LO. Solid and dashed lines
         represent nucleons and pions,
         respectively. Solid dots denote the leading-order vertices.  
\label{fig2} 
 }
  \end{center}
\end{figure}

Although the results for few-nucleon observables already look quite promising
 at N$^2$LO, it is mandatory to extend these studies to N$^3$LO as it has
 already been done
 for the 2N system. This would allow one to test the
convergence of the chiral
expansion and might shed some light on the
long-standing puzzles in the 3N continuum. The extension to N$^3$LO requires the
incorporation of the leading corrections to the 3NF 
which feed into five different topologies, see  Fig.~\ref{fig2}, and 
are currently being worked out. 
Interestingly, there are no contributions from higher-order vertices in the
effective Lagrangian (except $1/m$-corrections) so that the 3NF at N$^3$LO is
parameter-free. Notice further that the spin-space structure of the chiral 3NF
at N$^3$LO is much richer than the one of the currently available
3NF models. Partial results are already available for certain types of
contributions. In particular, one-loop and relativistic $1/m$-corrections to the
$2\pi$-exchange topology were considered in Refs.~\cite{Ishikawa:2007zz} and 
\cite{Coon:1986kq}, respectively. The results of Ref.~\cite{Ishikawa:2007zz}
based on the AV18 2NF \cite{Wiringa:1994wb} indicate that the
one-loop $2\pi$-exchange corrections have little effect on 3N scattering observables. 
   
\begin{figure}[tb]
\vskip 1 true cm
  \begin{center} 
\includegraphics[width=\textwidth,keepaspectratio,angle=0,clip]{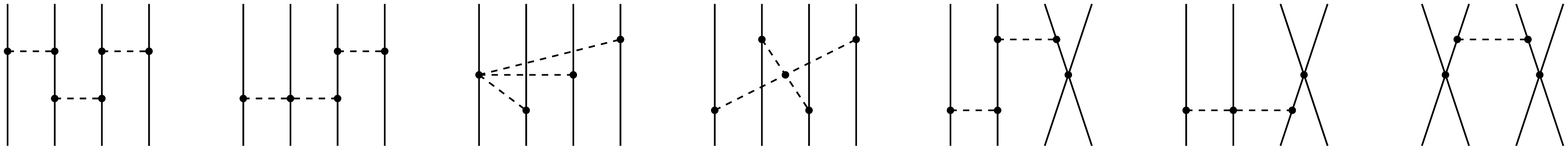}
    \caption{
         4N diagrams generating nonvanishing 4NF contributions at N$^3$LO.
         Diagrams which result from the interchange of the nucleon lines
         and/or application of the time reversal operation are not shown. For
         remaining notation see Fig.~\ref{fig2}.  
\label{fig3} 
 }
  \end{center}
\end{figure}

In addition to the 3NF corrections, four-nucleon force (4NF)
starts to contribute at N$^3$LO in the chiral expansion. It has been worked
out recently using the method of unitary transformation \cite{Epelbaum:2006eu}. 
The leading 4NF results from diagrams depicted in Fig.~\ref{fig3} where only  
graphs are shown which yield nonvanishing contributions. It is governed by the
exchange of pions and the lowest-order 
nucleon-nucleon contact interaction and includes effects due to the nonlinear  
pion-nucleon couplings and the pion self-interactions constrained by the chiral 
symmetry of QCD. The obtained 4NF is local and does not
contain any unknown parameters. The expectation values of the individual
4NF contributions obtained in the pioneering study of
Ref.~\cite{Rozpedzik:2006yi} are typically of the order of few hundreds of
keV, which agrees with estimations based on dimensional arguments. One also
observes strong cancellations between individual contributions.  
 
\section{Role of $\Delta$-excitations} 
\label{sec:delta} 
 
The $\Delta$-isobar is known to play an important role in hadronic and
nuclear physics due to its low excitation energy, $\Delta m \equiv m_\Delta -
m = 293$ MeV, and strong coupling to the $\pi N$ system. It can, therefore, be
argued that the explicit inclusion of the delta allows one to resum a certain
class of important contributions and thus leads to an improved convergence as
compared to the delta-less theory. Indeed, it is well known that e.g.~the LECs
$c_{3,4}$ accompanying the subleading $\pi \pi NN$ vertices are saturated by
the $\Delta$-isobar which leads to their rather large numerical values
\cite{Bernard:1996gq}. As a
consequence, the subleading $2\pi$-exchange 2NF turns out to be much
stronger than the leading one \cite{Kaiser:1997mw}. A similar tendency was also found 
for the $3 \pi$-exchange \cite{Kaiser:1999ff} and the charge-symmetry breaking
$2\pi$-exchange potentials \cite{Epelbaum:2005fd}. In EFT with explicit deltas, the
dominant portion of these large subleading contributions is shifted to
the lower chiral order leading to a more natural convergence pattern.  

\begin{figure}[b]
\vskip 1 true cm
  \begin{center} 
\includegraphics[width=9cm,keepaspectratio,angle=0,clip]{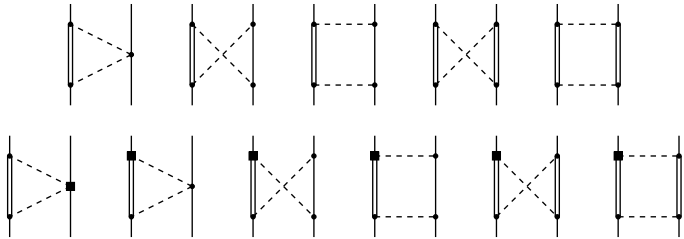}
    \caption{
         NLO (first raw) and N$^2$LO (second raw) contributions to the
         $2\pi$-exchange 2NF  with single
         and double $\Delta$ excitations. Double lines represent
         $\Delta$-isobars, filled squares denote subleading vertices.  For
         remaining notation see Fig.~\ref{fig2}.  
\label{fig4} 
 }
  \end{center}
\end{figure}

The explicit inclusion of the delta in chiral EFT can be achieved via the so-called
small scale expansion (SSE) \cite{Hemmert:1997ye} in which
$\Delta m$ is treated as an additional small parameter. The leading
contributions to the 2NF due to intermediate $\Delta$ excitations
arise at NLO from diagrams shown in the first raw of Fig.~\ref{fig4} and were
considered in \cite{Ordonez:1995rz,Kaiser:1998wa}. 
In \cite{Krebs:2007rh} we
worked out the subleading contributions which are generated at N$^2$LO by
diagrams depicted in the second raw of Fig.~\ref{fig4}. To obtain numerical
results, one needs to determine the combination of the the subleading $\pi N
\Delta$ LECs $b_3 + b_8$ which enters at this order and the LECs
$c_i$ whoes values  differ from the ones obtained in the delta-less theory. 
This was achieved by fitting the threshold coefficients of $\pi N$  
scattering calculated at second order in the SSE. As expected, we found that
the values of the LECs $c_{3,4}$ are reduced in magnitude in the theory with explicit
$\Delta$'s. Using the SU(4) (or large $N_c$) value for the leading $\pi N
\Delta$ LEC $h_A = 3g_A/(2\sqrt{2}) = 1.34$ with $g_A$ being the 
nucleon axial-vector coupling, we found $c_3 = -0.79$ GeV$^{-1}$ and $c_4 = 1.33$ GeV$^{-1}$
which has to be compared with $c_3 = -3.87$ GeV$^{-1}$ and $c_4 = 2.89$
GeV$^{-1}$ in the delta-less EFT.

\begin{figure}[tb]
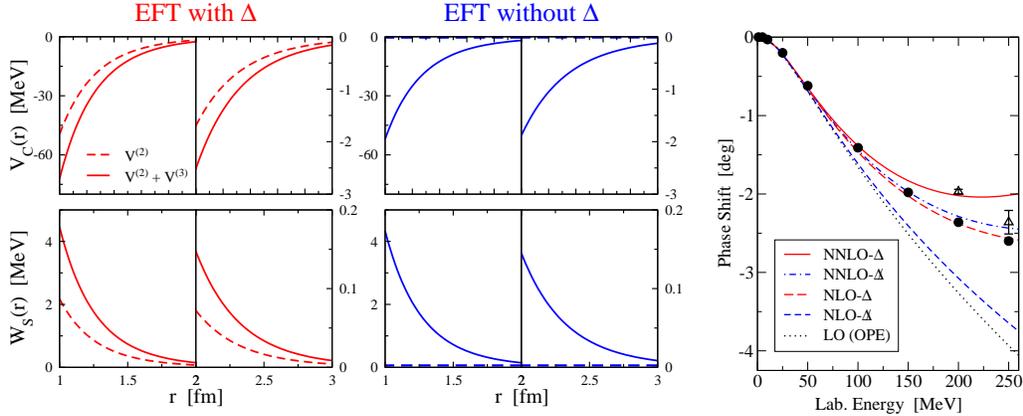

\vskip 1 true cm
  \begin{center} 

\begin{minipage}[c]{9cm}
\includegraphics[width=9cm,keepaspectratio,angle=0,clip]{vC_wS.eps}
\end{minipage}
\hfill
\begin{minipage}[c]{4.07cm}
\vskip 0.47 true cm
\includegraphics[width=4.07cm,keepaspectratio,angle=0,clip]{3f3.eps}
\end{minipage}
    \caption{
         Isoscalar central ($V_C (r)$) and isovector spin-spin ($W_S (r)$)
         potentials in $r$-space in EFT with (left panel) and without
         (middle panel) explicit $\Delta$'s. Right panel: $^3F_3$ NN phase shift in EFT with and without explicit
         $\Delta$'s calculated in first Born approximation. The filled circles
         (open triangles) depict the results
         from the Nijmegen multi-energy PWA \cite{Stoks:1993tb} (Virginia Tech single-energy
         PWA \cite{SAID}). In all cases, spectral-function regularization
         with the cutoff $\tilde \Lambda = 700$ MeV has been used.   
\label{fig5}
 }
  \end{center}
\end{figure}

In Fig.~\ref{fig5} we compare the isoscalar central and
isovector spin-spin $2\pi$-exchange contributions at NLO and N$^2$LO obtained in
EFT with and without explicit $\Delta$'s. 
Contrary to the delta-less theory where the whole contribution is generated at
N$^2$LO by subleading $2\pi$-exchange, the dominant contribution
in the delta-full theory results at NLO from the leading $2\pi$-exchange graphs. 
This much natural convergence pattern is also observed for peripheral phase
shifts, see Fig.~\ref{fig5} for an example. 

\section{Few nucleons on a lattice} 
\label{sec:lattice} 

\begin{figure}[tb]
\vskip 1 true cm
\begin{minipage}[c]{6.5cm}
\includegraphics[width=6.5cm,keepaspectratio,angle=0,clip]{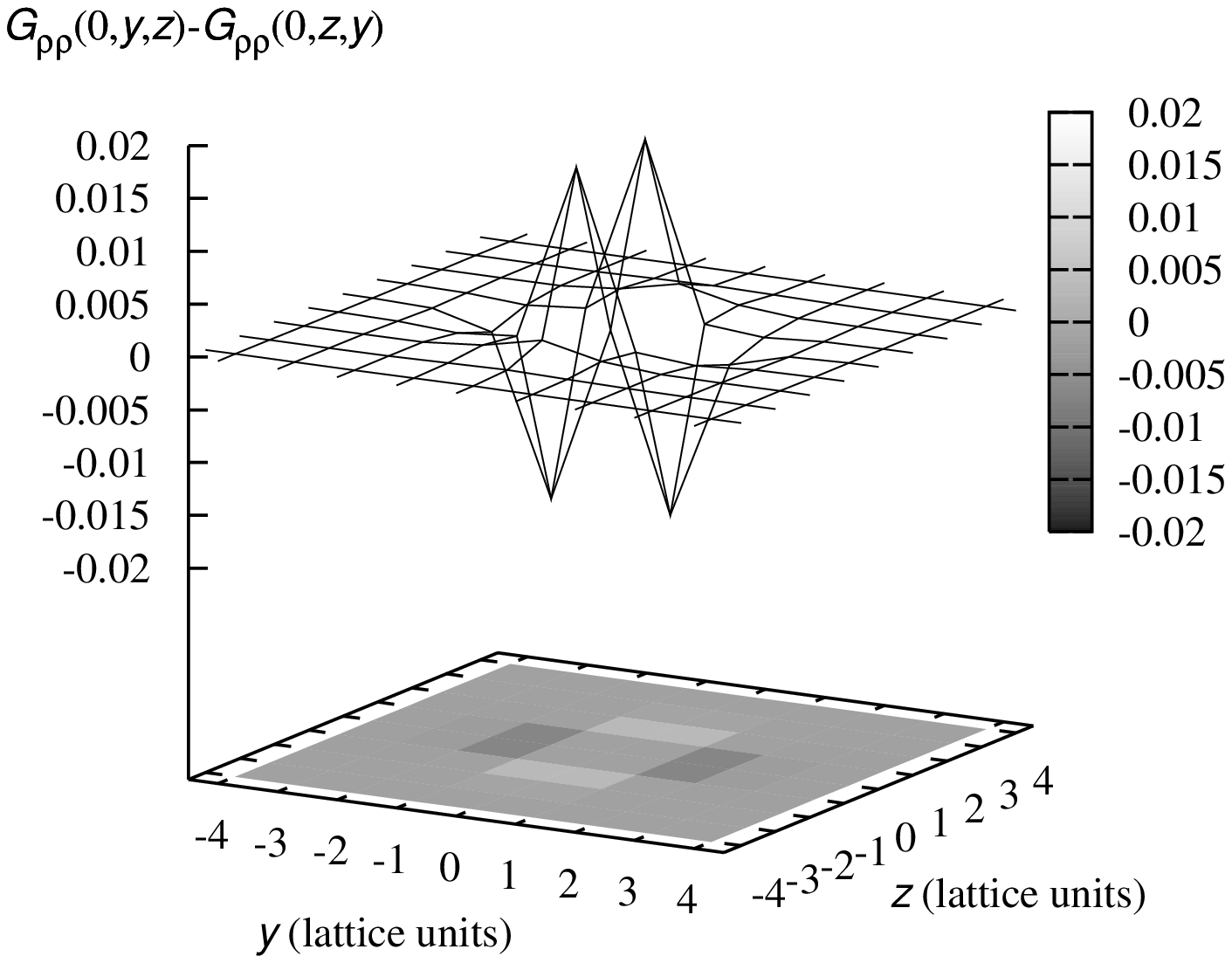}
\end{minipage}
\hfill
\begin{minipage}[c]{6cm}
\includegraphics[width=6cm,keepaspectratio,angle=0,clip]{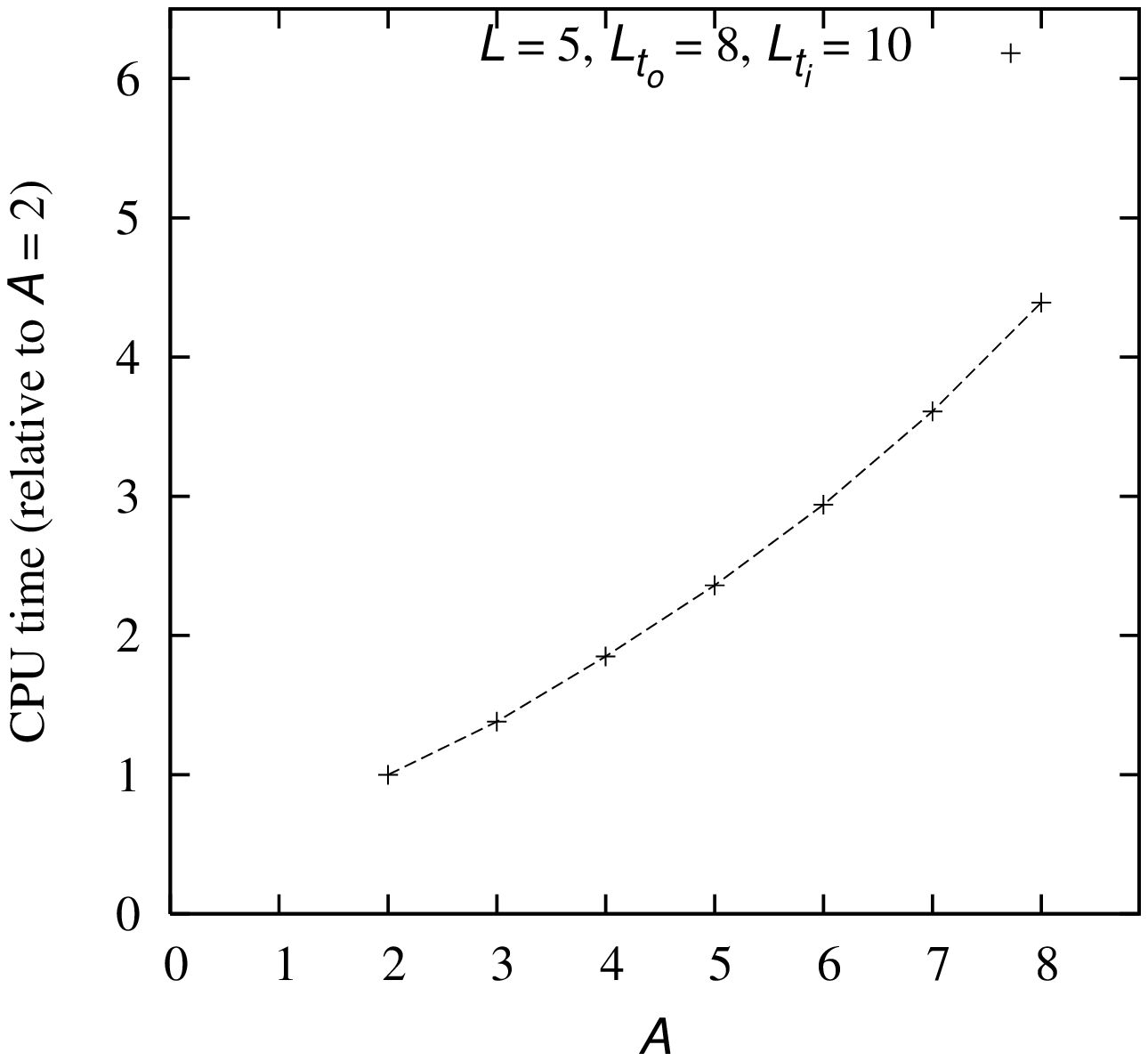}
\end{minipage}
\vskip -0.7 true cm
    \caption{
         Left panel: antisymmetric linear
         combination of the nucleon density correlations for the deuteron in the $yz$-plane. The deuteron
         spin points in the $+z$-direction. Right panel: CPU time versus the
         number  of nucleons, $A$, measured relative to the
         $A=2$ deuteron system.  
\label{fig7} 
 }
\end{figure}

The last topic I would like to address concerns nuclear lattice simulations
using chiral EFT \cite{Borasoy:2006qn}, see also \cite{Borasoy:2005yc} for a similar work
based on pionless EFT. In this framework, the effective Lagrangian is
formulated on a Euclidean lattice and the path integral is evaluated by Monte
Carlo sampling. Pions and nucleons are treated as point-like particles on
the lattice sites, and $\pi$ times the inverse lattice spacing sets the cutoff
scale in momentum space. By using hadronic degrees of freedom and
concentrating on low-energy physics, it is possible to probe larger volumes,
lower temperatures, and far larger numbers of nucleons than in lattice QCD.
In \cite{Borasoy:2006qn} simulations were carried out at lowest order 
in the chiral expansion. The sign problem is suppressed due to the SU(4)
positivity of the lattice path integral for any even number of nucleons and
the approximate SU(4) symmetry of the low-energy nuclear interactions. Having
fixed three unknown parameters entering the 2NF using the $^2$H binding energy,
the NN $^1S_0$ scattering length and the average S-wave effective
range,\footnote{The additional third parameter corresponds to higher-order
  contact interactions which were included to overcome a clustering
  instability, see    \cite{Borasoy:2006qn} for more details.} we computed 
various properties of light nuclei, see left panel in Fig.~\ref{fig7} for one example. 
The results for the deuteron properties and the obtained value for the triton
binding energy $E_{^3H} = -8.9(2)$ MeV are within 5\% of the
data while the $\alpha$-particle binding energy $E_{^4He} = -21.5(2)$ MeV is
about $25\%$ smaller in magnitude than the experimental value of $-28.296$ MeV. 
Encouraging results were also obtained for the computational scaling showing
for $A\leq8$ an approximately linear dependence on $A$, see right panel in 
Fig.~\ref{fig7}.

\section{Summary and outlook} 
\label{sec:summary} 

In this talk I discussed the structure of the nuclear force in few lowest
orders in the chiral expansion. In the 2N system, accurate description of
the low-energy data is achieved at N$^3$LO. The results for 3N
scattering and the properties of light nuclei are so far only available at
N$^2$LO. While most of the calculated 3N scattering observables are in a
reasonable agreement with the data, the theoretical uncertainty at this order
appears to be rather large. It is therefore mandatory to extend this studies
to N$^3$LO. The corrections to the 3NF which arise at this order are currently
being worked out. The leading 4NF which also contributes at N$^3$LO has
already been derived. 

In the second part of my talk I discussed the 2NF due to intermediate
$\Delta$-excitations at N$^2$LO in the SSE. It is demonstrated that the explicit
inclusion of the $\Delta$-isobar in chiral EFT allows to improve the
convergence of the low-momentum expansion. In the future, this work should be
extended to study the role of the $\Delta$-isobar in the 3NF and
isospin-breaking interactions. It would also be interesting to work out  
N$^3$LO contributions in the SSE. 

Finally, in the last part of my talk, I discussed the leading-order results
for the properties of light nuclei based on the lattice formulation of chiral
EFT. A generalization to higher orders in the chiral expansion and to
scattering observables, see \cite{Borasoy:2007vy} for the important step in
this direction, is in progress. 

\section*{Acknowledgements} 

I thank the organizers for the invitation and all my collaborators for sharing
their insight into the discussed topics.  This work was supported in parts by the 
Helmholtz Association under the contract number VH-NG-222.


\end{document}